\def\draftform{DRAFT}

\def\form{PRL}
\ifx\draftform\form
\documentclass[preprint,showpacs,preprintnumbers,superscriptaddress,amsmath,amssymb]{revtex4}
\else
\documentclass[prl,twocolumn,floatfix,showpacs,preprintnumbers,superscriptaddress,amsmath,amssymb]{revtex4}
\fi

\def\figwidth{\ifx\draftform\form 0.8\else0.45\fi}

\usepackage{graphicx}
\usepackage{epsfig}
\usepackage{bm}
\def\ltsim{\vbox {\hbox{\lower .8\baselineskip \hbox{$<$}} \break
                 \hbox{\lower 0.2\baselineskip \hbox{$\sim$}} } }
\def\k{{\bf k}}

\def\d{{\bf d}}

\def\r{{\bf r}}

\def\iv{{$I-V$ curve$\;$}}

\begin{document}
\bibliographystyle{apsrev}
\title{Current collapse in tunneling transport through benzene}

\author {M. H. Hettler} 
\affiliation{Forschungszentrum Karlsruhe, Institut f\"ur Nanotechnologie,
76021 Karlsruhe, Germany}

\author {W. Wenzel} 
\affiliation{Forschungszentrum Karlsruhe, Institut f\"ur Nanotechnologie,
76021 Karlsruhe, Germany}

\author {M. R. Wegewijs} 
\affiliation{Institut f\"ur Theoretische Physik A, RWTH Aachen, 52056
Aachen, Germany}

\author {H. Schoeller} 
\affiliation{Institut f\"ur Theoretische Physik A, RWTH Aachen, 52056
Aachen, Germany}

\date{\today}

\begin{abstract}
  We investigate the electrical transport through a system of
  benzene coupled to metal electrodes by electron tunneling.  Using
  electronic structure calculations, a semi--quantitative model for
  the $\pi$ electrons of the benzene is derived that includes general
  two-body interactions. After exact diagonalization of the benzene
  model the transport is computed using perturbation theory for weak
  electrode-benzene coupling (golden rule approximation). We include
  the effect of an applied electric field on the molecular states, as
  well as radiative relaxation. We predict a current collapse and strong
  negative differential conductance due to a ``blocking'' state when
  the electrode is coupled to the para-position of benzene. In
  contrast, for coupling to the meta-position, a series of steps in the $I-V$ curve is
  found.
\end{abstract}

\maketitle

{\em Introduction.} Single molecule electronic devices
offer exciting perspectives for further minituarization of
electronic circuits with a potentially large impact in applications.
Several experiments have demonstrated the possibility to attach
individual molecules to leads and to measure the electrical transport.
~\cite{reed-etal,kergueris-etal,reichert-etal,platin-complex} 
In contrast to single electron transistors (SETs) based on quantum
dots~\cite{sohn-etal} the electronic structure of molecular devices can be
chemically 'designed' for specific applications. When the molecule is
coupled {\em weakly} to the electrodes, i.e. via electron tunneling,
charging effects, semi-classically determined by the small capacitance
of the molecule, become important. The interplay of charging effects
with the specific structure of the molecular orbitals leads to
nontrivial current voltage ($I-V$)
characteristics~\cite{chen-etal,epl_paper}. Very recent experiments
demonstrated both Coulomb blockade and the Kondo effect in three
terminal transport through a single molecular level~\cite{cocomplex,v2complex}.

Using benzene as a prototypical example we investigate novel effects
that arise when transport through several competing electronic
configurations becomes possible. We derive a semi-quantitative model
for the conducting many-body states of the system from electronic
structure calculations. For weak coupling to the electrode we compute
transport within the golden rule approximation (sequential tunneling)
and include screening of the applied electric field as well as
radiative transitions between the electronic states of the molecule.
We predict a current collapse in the $I-V$ curve and strong negative
differential conductance (NDC) due to the occurrence of a ``blocking''
state when the molecule is coupled to the electrode at the
para-position.  For coupling at the meta-position, the $I-V$ curve displays a
series of steps, but no NDC.  We demonstrate how the specific {\it
spatial} structure of the molecular orbitals qualitatively determines
electronic transport.  Finally, we discuss the limits of the model and
the impact of disorder and symmetry breaking effects likely
encountered in an experimental realization.

{\em The Model.} To perform transport calculations in the weakly
coupled regime, we extract an effective model from electronic
structure calculations of the molecule. For benzene transport is
assumed to be dominated by the $\pi$-electron system, but
generalizations are straightforward. We first perform Hartree-Fock
calculations in a suitable basis and then transform the Hamiltonian to
the molecular orbital basis. We then integrate out the $\sigma$
electrons of the system arriving at an effective {\em interacting}
model Hamiltonian for the $\pi$ electrons of the system
in the presence of the atomic cores and the ``frozen'' density of the
$\sigma$ electrons:
\begin{equation}
{\cal H_\pi} = \sum_{ij\sigma} 
\epsilon_{ij} c^\dagger_{\sigma,i} c_{\sigma,j}
+ \sum_{ijkl\sigma\sigma'} U_{ijkl} c^\dagger_{\sigma,i} c^\dagger_{\sigma',j}
c_{\sigma,k} c_{\sigma,l} \label{eq:heff}
\end{equation}
The second quantized operators $c^\dagger_{\sigma,i}, c_{\sigma,i}$
create/destroy electrons of spin $\sigma$ in orthogonalized
Wannier-like orbitals $\Phi_i$ centered at the carbon atoms. While this model
neglects $\sigma$-$\pi$ mixing for certain excited states of the
molecule~\cite{roos92,hirao96}, its parameters account for the
detailed electronic structure of the molecule. For the current work,
we compute the model parameters only once, using an augmented
double-zeta quality atomic natural orbital (ANO) basis set that was
truncated to contain only one 2p-shell for each carbon. When applying
a bias over the molecule this neglects the higher order effect of
field screening by $\sigma$ electrons and its impact on the $\pi$
electrons. Note that in our approach 
there is no over-counting of interaction terms like 
in ``interaction enhanced'' 
density functional treatments in solid state physics~\cite{steiner,lda+dmft}.

We find that the low-energy spectrum of benzene obtained from the
diagonalization of the effective Hamilton operator eq.~\ref{eq:heff}
compares favorably with the spectrum directly obtained from accurate
multi-reference configuration interaction
calculations~\cite{hirao96,stampfuss99}. The remaining differences can
be understood by the lack of $\sigma$-$\pi$ mixing and do not
qualitatively affect the transport properties. The restriction to a
pure $\pi$-electron system is more severe for the charged states, as
discussed in some detail below.

To account for the effect of an external  bias potential $V^{ext}$ 
on the electrons we include a term 
\begin{equation}
{\cal H}_{bias} = e \sum_{ij\sigma\sigma'} 
V^{ext}_{ij} c^\dagger_{\sigma,i} c_{\sigma,j} \label{eq:hbias}
\end{equation}
with $V^{ext}_{ij} = \int d\r \Phi_i (\r) V^{ext} (\hat{\r}) \Phi_j
(\r) $ and $V^{ext} (\hat{\r}) = (V_L+V_R)/2 - V_{bias} (\hat{\r}/L)$
in the Hamiltonian. $V_{L,R}$ is the chemical potential in the
left/right electrode, distances are measured from the center of the
molecule. The matrix elements $V^{ext}_{ij}$ also result from the
electronic structure calculation described above. We consider a bias
$V_{bias}=V_L-V_R$ aligned with the transport direction ($x$-direction).
The parameter $L$ indicates the length over which the
external bias falls off (we choose $L=0.4nm$).

{\em Diagonalization and Transport Calculation.}  To compute the
transport properties of the system in the weak coupling limit we
diagonalize the Hamiltonian eq.~\ref{eq:heff} in the appropriate
charge/spin/symmetry sectors.  
For the current model, we have one "single-particle" state (Wannier
state $\Phi_i$) per carbon site $i$, resulting in a total
of $4^N=4096$ (with $N=6$ sites) many-body states $| s \rangle$. For the
present model, charge, spin and symmetry adaptation reduce the size of
effective Hamiltonian matrices sufficiently to permit their
diagonalization with standard linear algebra packages. Care is needed
to handle degeneracies properly. The energy degenerate states need to
be additionally diagonalized to obtain states with integer
(half-integer) total spin for states with even (odd) number of
electrons. This also helps for the transport calculation, since the
transition matrix elements can be summed over the magnetic quantum
numbers, i.e. we only need to consider one representative of the
degenerate spaces (we do not consider application of a magnetic field
here).

After diagonalization of the Hamiltonian we have the many-body
eigenstates $| s \rangle$ with the corresponding energies $E_s$ and
their total spin $S_s$. We use a Master equation
approach~\cite{epl_paper} for the occupation probabilities $P_s$ in a
stationary state. The transition rates $\Sigma_{ss'}$ from state $s'$
to $s$ are computed in perturbation theory using the golden rule. The
"perturbation" is the coupling of the molecule to the leads
\begin{equation}
H_{mol-leads} =   (\frac{\Gamma}{2\pi\rho_e})^{1/2}
\sum_{\k\sigma\alpha l}\left(
c^{\dagger}_{l\sigma}a_{\k\sigma\alpha} + h.c. \right)\;,
\end{equation}
and (optionally) the coupling to electromagnetic fields (photons).
$\Gamma$ is the coupling strength (in units of energy) of leads to the
benzene and $\rho_e$ is the density of states of the electrons in the
electrode (assumed constant). The operators
$a_{\k\sigma\alpha}$ and their hermitian 
conjugates destroy/create electrons with momentum $\k$ and spin $\sigma$ in
electrode $\alpha=$ left/right. 
For simplicity, we assume that
tunneling is only possible through two ``contact'' carbon atoms which we
choose to be at the 1 and 4 (para) positions unless noted otherwise.

As we do not consider the leads microscopically, the coupling of
molecule states $| s \rangle$ is determined by the overall coupling
strength $\Gamma$ and the relative wave function amplitude of the
state $| s \rangle$ at the coupling carbon site $l$.  For the transition
rates we have $\Sigma_{ss'}=(\sum_{\alpha,p=\pm}\Sigma_{ss'}^{\alpha
p}) +\Sigma^d_{s,s'}$ where $\Sigma_{ss'}^{\alpha p}$ is the tunneling
rate to/from electrode $\alpha$ for creation ($p=+$) or destruction
($p=-$) of an electron on the molecule. We have
\begin{equation}
\Sigma_{ss'}^{\alpha +}=\Gamma f_\alpha(E_s-E_{s'})
\sum_\sigma |\,\sum_l
<s|c^\dagger_{l\sigma}|s'>\,|^2\;,
\end{equation}
and a corresponding equation for $\Sigma_{s's}^{\alpha -}$ by
replacing $f_\alpha\rightarrow 1-f_\alpha$, where $f_\alpha$ is the
Fermi function.  When including relaxation by radiative transitions,
we use the dipole approximation with dipole transition moments
$\d_{i,j}= \int d\r \Phi_i (\r) \hat{\r} \Phi_j (\r) $ obtained from
the electronic structure calculations. The corresponding transition
rates are
$$\Sigma^d_{ss'}=\frac{4 e^2}{3 \hbar^3 c^3}
(E_s-E_{s'})^3 N_{b}(E_s-E_{s'})
\, | <s|\d|s'>|^2\;
$$
where $N_{b}(E)$ denotes the equilibrium Bose function.
Note that for emission $E_s-E_{s'}$ is negative, and 
$N_{b}(-|E|) = -(1+N_{b}(|E|))$.

The total transition matrix $\Sigma_{ss'}$ consists  of blocks
connecting $N$ and $N\pm1$ electron states (tunneling processes)
and blocks from  the radiative transitions that do not 
change the electron number on the molecule. 
Taking only one member of the subspace of spin and energy degenerate
states into account, the rank $r$ of the
transition matrix is 1716.
The stationarity condition $\dot{P_s}= 0$ can be written as 
$\sum_{s'} A_{ss'}  P_{s'} = 0$
with the matrix $A_{ss'} =\Sigma_{ss'} -  \sum_{s''} \Sigma_{s''s} 
\delta_{ss'}$.
This implies that $A_{ss'}$ has an eigenvector with zero eigenvalue,
which is the wanted solution for $P_s$.
Rather than computing this eigenvector by brute force, to speed up the 
calculation  we make use of  $\sum_{s} P_s =1$ to eliminate one
row/column, thus reformulating  the eigenvector problem  into one of 
solving an inhomogeneous linear system of rank $r-1$.
Eventually, the current in the left and right electrode is calculated via 
$ I_{\alpha}= e \sum_{s,s'} (\Sigma^{\alpha +}_{ss'} P_{s'} -
\Sigma^{\alpha -}_{s's} P_s)\;.$

In this paper, we consider symmetric bias only. Applying an asymmetric
bias would effectively imply different coupling to the left and right
leads and has been previously discussed~\cite{epl_paper}. Finally, we
note that we must have $\Gamma \ll k_b T \approx 40meV$ for
perturbation theory to apply at room temperature T.

\begin{figure}[t]
\begin{center}
\includegraphics[width=8 cm]{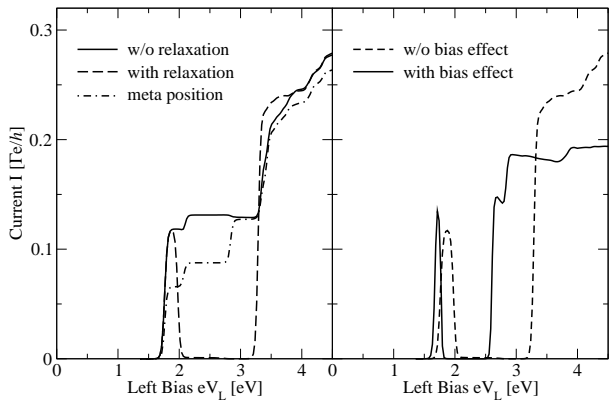}
\end{center}
\caption {$I-V$ characteristics for symmetric bias.
Left Panel: \iv without (solid line) and with radiative relaxation
(dashed line) without inclusion of bias effect.  The current collapses
at a bias when the antisymmetric anion state can become occupied by
radiative relaxation from an excited anion state. For coupling of the
right electrode at the meta position (dash-dotted line) 
the ``blocking state'' has finite tunnel coupling to both electrodes
and no current collapse is observed. 
Right Panel: With inclusion of the bias effect (solid line) the onset of
current is generally shifted to lower bias, but the current collapse
remains and additional weak NDC occurs at larger bias.}
\label{gen_ndc}
\end{figure}

{\em Results:} To elucidate the impact of various effects on the
current we have performed transport calculations 
with and without radiative transitions (relaxation) and with and without 
effect of the applied bias. 

On the left panel of Fig.~\ref{gen_ndc} we present the \iv obtained without
the effect of the applied field (zero bias electronic
structure). Without radiative relaxation (solid line) the \iv consists
of a series of steps of which only a few are well resolved on this
scale. The first step is associated with the population of the first
$\pi^*$ orbital of molecule (molecular charge $= -e$), an electron hops
onto the lowest available level and then hops off again. At slightly
larger bias, a transition of the anion to the first excited state of
the neutral molecule becomes possible, resulting in a slight increase
of the current. If the bias is sufficiently large this excited state
may now accept another electron to populate higher excited states of
the anion or low-lying states of the di-anion, resulting in a rapidly
growing cascade of transitions between literally hundreds of states of
the system. In the model considered here the growth of this cascade
leads to quasi-ohmic behavior above 3.6 eV.  In our calculation the
first states of the di-anion become occupied at about 4.5 eV.

\begin{figure}
\begin{center}
\includegraphics[width=6 cm]{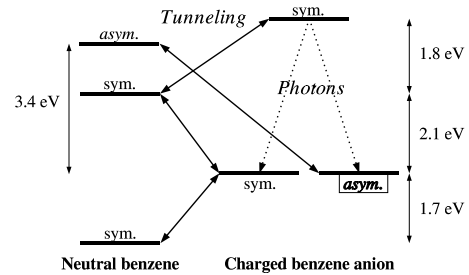}
\end{center}
\caption {Sketch of the energetics and symmetry of the relevant neutral
and anion states. At a left bias of 2.1 eV the antisymmetric 
blocking state becomes occupied via
radiative relaxation. At 3.4 eV electrons can escape the blocking
state via tunneling to the antisymmetric state of neutral benzene.
}
\label{energy_sketch}
\end{figure}

The inclusion of radiative transitions has a dramatic effect
on the \iv (dashed line). We observe a collapse of the current over a
substantial range of the applied bias (2.1-3.4 eV). The reason for
this collapse is the population of a ``blocking'' state in the cascade
of transitions that becomes possible when exited states of the neutral
molecule and anion become accessible. Above approximately 2.1 eV bias
an excited state of the anion at about 5.6 eV (see Fig.~\ref{energy_sketch}), becomes
partially populated in the transport cascade. This state can decay by
photon emission to either a symmetric or an antisymmetric many body
state (with respect to the plane through the transport axis and perpendicular to the
molecular plane) of the anion.  In the bias range of the current collapse this state
cannot decay by coupling to the leads, because the lowest neutral
states are symmetric (see Fig.~\ref{energy_sketch}) and the tunneling
preserves the symmetry. Since there are no further radiative
transitions possible on the molecule, the rate equations contain no
draining term from this state. As a result, in the stationary state, the
probability of occupying the ``blocking'' state is unity and the current
ceases to flow. At a larger bias (3.4 eV), the first escape channel
opens, and the system can decay to the first antisymmetric excited
state of the neutral molecule, which can then decay further by
photon emission.

The above calculations simplify the illustration of the mechanism of
NDC, because the energies of the participating states are independent
of the applied bias. In the physical system the various states couple
differently to the applied field. The solid line in the right panel of
Fig.~\ref{gen_ndc} shows the \iv of the benzene with radiative relaxation and with
the effect of the external potential applied parallel to the transport
axis. We note a shift of the first step in the $I-V$ curve, which is due to
differential screening effects between the ground state of the neutral
molecule and the anion. At larger bias, the differential effects on
all of the states in the cascade result in a significant
renormalization of the $I-V$ curve. However, the current collapse still
occurs, although the voltage window is reduced. The onset of
quasi-ohmic behaviors occurs at higher voltage and a series of weak
NDC effects occurs due to redistribution of occupation probabilities
in favor of states with slightly smaller ``transmission''.

If we couple the right electrode to the meta position of benzene (left
panel, dash-dotted curve) we find no current collapse with or without
radiative relaxation or bias effect. This can be readily understood by
the fact that at the meta position the wave function of the formerly
blocking state is non-vanishing and electrons can tunnel out to the
right electrode.  Consequently, we observe a series of current steps
similar (but not identical) to the case of coupling to the para
position without relaxation.

{\em Discussion:} In the above we have described transport in an
idealized, though semi-quantitative, $\pi$-electron model of weakly
coupled benzene and found a dramatic suppression of the current in a
finite voltage window. The mechanism for this effect is the occupation
of a so-called blocking state of the molecule,
which cannot decay for symmetry reasons. In the following we will summarily
discuss its stability under the influence of a number of effects not
considered in our calculation, but likely to be  present in a experimental
realization. Details will be discussed in a forthcoming
publication~\cite{rydberg}. 

(i) {\em Transport through the $\sigma$-system: } The ionization
potential of benzene is rather high, for this reason no cationic
states (with either $\sigma$ or $\pi$ holes) are relevant for the bias
range considered here and attainable in experiment. However, for the
anion, low lying Rydberg states are predicted in quantum chemical
calculations with augmented basis sets containing diffuse
functions. In these states the additional particle occupies an
electron cloud of $\sigma$ symmetry that is smeared out over the
entire molecule and has significant charge density in the center of
the ring. Including these states, we expect the onset of current in
the $I-V$ curve, i.e. the first current step, will be due to tunneling into
the lowest Rydberg state, at about 1 eV. Nevertheless we believe that
the NDC effect will be preserved.  The current collapse is due to the
competition between a slow process (cascade filling the blocking
level) with an even slower one (decay of the blocking level). 
NDC occurs when probability density is redirected from ``conducting''
many-body states with large tunnel coupling to non-conducting
many-body states even as the voltage increases. Note that the
existence of a single ``blocking path'' is sufficient for NDC,
irrespective of the number of competing non-blocking cascades.

The addition of Rydberg states below the blocking $\pi$ state of the
anion permit its decay to symmetric or antisymmetric Rydberg states of
the anion. Note that the spectrum of the neutral molecule is
unaffected --- no additional low-lying configurations must be
considered. Therefore a new radiative cascade on the anion ultimately
leads to a symmetric Rydberg state, which can couple to the ground
state configuration of the neutral molecule. The amplitude for this
relaxation, however, is slow, because the transition dipole moment
from the blocking ($\pi$) configuration to the Rydberg ($\sigma$)
configuration is small~\cite{rydberg} (involving a dipole transition
moment perpendicular to the benzene plane).

{\em (ii) Symmetry breaking:} In most experimental realizations for
two terminal transport, with the possible exception of STM
contacts~\cite{stmscience01}, the alignment of the molecule with
respect to the electrodes is uncontrolled. It is therefore likely that
the applied bias is not aligned with the transport direction. In this
case the degenerate states of the anion mix, and transport is no
longer completely blocked. However, according to the rate argument
presented above, it is still impeded, and a reduced NDC remains
observable. The same consideration holds true for slow vibrational
modes of the molecule, like breathing and buckling~\cite{diventra}.
Fast vibrations can provide radiationless relaxation and should be
considered at the quantum level. They can be particularly important if
they generate transition matrix elements that allow for dipole
forbidden electronic transitions in the undistorted system.

{\em Summary:} In weakly coupled molecules the concept of the
``blocking'' state suggests new opportunities for the design of single
molecule electronic devices. In this investigation we have
demonstrated how transport through several competing many-body states
can result in strong NDC and current collapse over a significant
voltage window.  Our calculations predict the existence of this
effect for a realistic molecule. Note that for $V_{bias}$ in the
vicinity of the current peak, the application of a gate voltage $V_g$
may induce a large asymmetric change in the current, which would first
rise and then drop again upon scanning of $V_g$ through a certain
range. This behavior could be useful in various electronic devices.
	
{\em Acknowledgments.}	 
One of us (M.R.W.) would like to thank Philip Stampfu{\ss} for assistance
with the electronic structure calculations. WW acknowledges support
of the DFG (grant WE 1863/10-1), the von-Neumann-Institute for
Computing and the HLRZ Karlsruhe.

\end{document}